# Effect of Colorlessness Condition on Phase Transition from Hadronic Gas to Partonic Plasma

M. A.A. Ahmed[a,b,c], S. Cherif[c,d], M. Ladrem[e,c], H. Zainuddin[b,f] and N. M. Shah[b,f]

[a]Physics Department, Faculty of Education, Arts and Sciences, Taiz University, Taiz, Yemen
[b]Institute for Mathematical Research, Universiti Putra Malaysia, 43400 Serdang Selangor
[c]Laboratoire de Physique et Mathematiques Appliquées, ENS-Vieux Kouba, Algiers, Algeria
[d]Sciences and technologies Departement, Ghardaia University, Algeria.
[e]Physics Department, Faculty of Science, Taibah University, Al Madinah Al Munawwarah, KSA.
[f]Department of Physics, Faculty of Science, Universiti Putra Malaysia, 43400 Serdang Selangor
[*]mohammed_h7@yahoo.com
mladrem@yahoo.fr

**Abstract.** One of the most important phase transition in physics is the Deconfinement Phase Transition in thermal Quantum ChromoDynamics. Due to the confinement property, we study the effect of colorlessness condition during the Deconfinement Phase Transition from a Hadronic Gas to a Quark-Gluon Plasma. We investigate the behavior of some thermodynamical quantities of the system such as the energy density and the pressure, the colorlessness condition and without colorlessness.

## INTRODUCTION

It is well-known that hadrons have quark substructure and the strong interaction between them are mediated by charged gluons and is described by a gauge theory known as Quantum ChromoDynamics (**QCD**). This theory has two remarkable properties:-
- Asymptotic freedom at high-energy and/or high temperature.
- Confinement at low energy and/or low temperature.

The confinement of color charge prohibited us to observe quarks as free particles and that all hadrons have zero color. We can then imagine that any compression and/or heating of the hadronic matter can induce dissolution of hadrons into a free Quark Gluon Plasma (QGP). This is commonly called the Deconfinement Phase Transition (DPT). The interest in the DPT is kept alive due to the need to explain the confinement phenomenon and to know the properties of the QGP with its evolution in the early universe since the big bang.

Experimentally, the predicted QGP state at high temperatures and densities, has been realized in the laboratory [1-3]. The resulting system does not quite behave as a quasi-ideal state of free quarks and gluons, but, rather, as an almost perfect dense fluid [4-5]. In this work, we try to analyze the behavior of the colorlessness condition effects by calculating some thermodynamical quantities in the vicinity of the transition point, in particular analyzing their behavior as a function of the temperature. Quark confinement in QCD cannot be treated perturbatively and has not been shown analytically, until now, for large distance or low density. However, the confinement has been almost confirmed numerically by lattice calculations. In lattice QCD, time-consuming calculations have shown that the concepts of quark confinement and asymptotic freedom are inherent in the model of strong interaction of matter at the sub-hadronic level, in addition, leading directly to a great deal of interest in the associated phase transition. The hypothesis of the colorless states (Singlet of **SU(3)** colour group) of a partonic system as a subsidiary requirement is assumed in describing the global thermodynamical equilibrium state.

Therefore, part of this work focuses on quantities when we are taking the colorless condition into account of the total partition function for the partonic plasma with the remaining large part of non-perturbative effect of QCD interaction.

## STATISTICAL STUDY OF THE SYSTEM

- **Partition Function:**

We take the Hadronic Gas (HG) as an ideal relativistic gas of massless pions and the QGP phase is just a free partonic system. For this purpose, we consider a simple Phase Coexistence Model (PCM) used in [6], in which the mixed phase system has a finite volume $V = V_{HG} + V_{QGP}$. The parameter h represents the fraction of volume occupied by the HG: $V_{HG}=h\,V$ and can be considered as an order parameter for the QCD-DPT. Assuming non-interaction between the phases, the total partition function of the system can be written as:

$$Z_{sys}(T,V,\mu_1,\mu_1,\dots) = Z_{HG}(T,V,\mu_1,\mu_1,\dots)Z_{QGP}(T,V,\mu_1,\mu_1,\dots)Z_{VA}(T,V,\mu_1,\mu_1,\dots)\dots(1)$$

To specify the confinement of quarks and gluons, the real vacuum pressure exerted on the perturbative vacuum B of the bag model should be taken. The mean value of any thermodynamical quantity of the system defined in [6], can then be calculated by:

$$\langle A(T,V,\mu)\rangle = \frac{\int_0^1 A(T,V,\mu)Z_{sys}(T,V,\mu)dh}{\int_0^1 Z_{sys}(T,V,\mu)dh}\dots(2)$$

where $A(h, T, \mu, V)$ is the total thermodynamical quantity in the state $h$. In the case of an extensive quantity, this is given by:

$$A(T,V,\mu) = A_{HG}(T,V\,h,\mu) + A_{QGP}(T,V\,(1-h),\mu)\dots(3)$$

and in the case of an intensive quantity, by:

$$A(T,V,\mu) = h\,A_{HG}(T,V\,h,\mu) + (1-h)\,A_{QGP}(T,V\,(1-h),\mu)\dots(4)$$

where $A_{QGP}$ and $A_{HG}$ the individual relative contributions from $QGP$ and $HG$ phases, respectively

- **The Partition function without colorlessness condition:**

The partition function of the grand canonical ensemble of the Fermi-Dirac and Bose-Einstein is given by:

$$Z_{B,F}^{GC}(T,V,\mu) = \prod_i \left(1+\sigma e^{-\beta(\varepsilon_i-\mu)}\right)^\sigma \dots(5)$$

where σ is a parameter with σ =+1 for a fermi gas and σ =-1 for a bose gas; $\beta = \frac{1}{k_B T}$ with T as the temperature and $k_B$ as the Boltzman constant. In the continuous case Eq.(5) can be written as (see [7])

$$\ln Z_{B,F}^{GC}(T,V,\mu) = \frac{\delta\,V}{2\pi^2}\sigma\int_0^\infty p^2 dp\left(1+\sigma e^{-\beta(\varepsilon_i-\mu)}\right)\dots(6)$$

where p is the momentum and $\delta$ the energy-level degeneracy. The partition function of the grand canonical ensemble of the of the relativistic quantum gas, with particle mass $m$ and chemical potential $\mu$ in volume $V$ and temperature $T$, can be calculated for quark and anti-quark system as (see [8]).

$$\ln Z_{Q\bar{Q}}^{NC}(T,V,\mu) = \frac{V_Q}{6\pi^2}\beta\left(\frac{1}{\hbar c}\right)^3\sum_{i=1}\delta_i^{q\bar{q}}\left(\frac{\mu^4}{2}+\frac{7\pi^4}{30}\frac{1}{\beta^4}+\pi^2\frac{\mu^2}{\beta^2}\right)$$

$$= V_Q T^3\left(\frac{1}{2\pi^2}\left(\frac{\mu}{T}\right)^4+\left(\frac{\mu}{T}\right)^2+\frac{7}{30}\pi^2\right)\quad\dots(7)$$

where $\delta = 4$, $V_Q$ volume of the Quarks, $\hbar$ the Planck constant, c the speed of light.

The gluon partition function is simply, given by

$$\ln Z_{G\bar{G}}^{NC}(T,V,\mu) = \frac{V_G}{6\pi^2}\beta\left(\frac{1}{\hbar c}\right)^3\sum_{i=1}\delta_i^{g\bar{g}}\left(-\frac{(\beta\mu)^4}{4\beta^4}-\frac{6}{\beta^4}\alpha(\mu)+\frac{12}{\beta^4}\left[\frac{\pi^4}{90}+\frac{\pi^2}{6}(\beta\mu)^2\right]\right)$$

$$= V_{G\bar{G}}T^3\left(\frac{1}{3}\left[16\left(\frac{\mu}{T}\right)^2-\frac{2}{\pi^2}\left(\frac{\mu}{T}\right)^4\right]-\frac{16}{\pi^2}\alpha(\mu)+\frac{16}{45}\pi^2\right)\quad\dots(8)$$

where $\alpha(\mu) = \sum_{p=1}^\infty\frac{Exp^{-p(\mu\beta)}}{p^4}\xrightarrow{\mu=0}\sum_{p=1}^\infty\frac{1}{p^4}=\frac{\pi^4}{90}$. Concerning the partition function for a pionic gas, this can be calculated from the bosonic partition function and is given by,

$$\ln Z_\pi^{NC}(T,V,\mu) = \frac{V_\pi}{6\pi^2}\beta\left(\frac{1}{\hbar c}\right)^3\sum_{i=1}\delta_i^\pi\left(-\frac{\mu^4}{4}-\frac{6}{\beta^4}\alpha(\mu)+\frac{12}{\beta^4}\left[\frac{\pi^4}{90}+\frac{\pi^2}{6}(\beta\mu)^2\right]\right)$$

$$= V_\pi T^3\left(\left[\left(\frac{\mu}{T}\right)^2-\frac{1}{8\pi^2}\left(\frac{\mu}{T}\right)^4\right]-\frac{3}{\pi^2}\alpha(\mu)+\frac{\pi^2}{15}\right)\dots(9)$$

Given $B$ is the bag constant accounting for the real vacuum pressure exerted on the perturbative vacuum, we have

$$lnZ_{VA}(T,V,\mu) = -\frac{BV}{T} \quad \ldots (10)$$

For a *QGP* consisting of gluons and two flavors of massless quarks (*up, down*) with a chemical potential $\mu$ in the bag model the total partition function will be given as follow:

$$Z_{TOTAL}^{NC}(T,V,\mu) = \left[Z_{Q\bar{Q}}^{NC} Z_{GG}^{NC} Z_{\pi}^{NC} Z_{VA}^{NC}\right](T,V,\mu)$$

$$= Exp^{V_\pi T^3\left(\left[\left(\frac{\mu}{T}\right)^2 - \frac{1}{8\pi^2}\left(\frac{\mu}{T}\right)^4\right] - \frac{3}{\pi^2}\alpha(\mu) + \frac{\pi^2}{15}\right) + V_{QGP}T^3\left(\frac{19}{3}\left(\frac{\mu}{T}\right)^2 + \frac{1}{6\pi^2}\left(\frac{\mu}{T}\right)^4 - \frac{16}{\pi^2}\alpha(\mu) + \frac{53}{90}\pi^2 + \frac{B}{T^4}\right)} \quad \ldots (11)$$

When $\mu = 0$, we have

$$lnZ_{TOTAL}^{NC}(T,V) = V T^3\left[\left(\frac{37}{90}\pi^2 - \frac{B}{T^4}\right) + h\left(-\frac{17}{45}\pi^2 + \frac{B}{T^4}\right)\right] \quad \ldots (12)$$

- **The Partition function colorlessness condition:**

In this part, we introduce the colorlessness condition under the projection method [9]. We consider the general case that the quarks have color-isospin with the color-charges emerging as degrees of freedom, but on another hand the hadrons have no color. So in calculating of the partition function of the *QGP* phase we take the trace for all physical color states and use the projection method. This method was based on the internal symmetry group given by [10].

$$g\left(\varphi,\psi,\frac{\mu}{T}\right) = \frac{\pi^2}{12}\left(\frac{21}{30}d_Q + \frac{16}{15}d_G\right) + \frac{\pi^2}{12}\frac{d_Q}{2}\sum_{q=r,b,r}\left\{-1 + \left(\frac{\left(\Theta_q - i\left(\frac{\mu}{T}\right)\right)^2}{\pi^2} - 1\right)^2\right\}$$

$$-\frac{\pi^2}{12}\frac{d_G}{2}\sum_{g=1}^{4}\left(\frac{(\Theta_g - \pi)^2}{\pi^2} - 1\right)^2 \quad \ldots (13)$$

$\mu$ is a chemical potential, T temperature, $d_Q = 2N_f$, $N_f$ the number of flavors $d_G = 2$ and degeneracy factors of quarks and gluons respectively, $\Theta_q$(q=r,b,g) the angles determined by the eigenvalues of the color charge operators and $\Theta_g$(g=1,…,4).

The colorless partition function is given by:

$$Z^{CS}(T,V_{QGP},\mu) = \frac{4}{9\pi^2}Exp^{-\frac{B}{T}V_{QGP}}\int\int_{-\pi}^{\pi}d\varphi d\psi\, M(\varphi,\psi)\, Exp^{V_{QGP}T^3\, g\left(\varphi,\psi,\frac{\mu}{T}\right)} \quad \ldots (14)$$

Let us derive in the following such an approximate color-singlet partition function at zero chemical potential $\mu = 0$. In this case, we have:

$$g(\varphi,\psi,\mu=0) = \frac{37}{90}\pi^2 - \frac{2}{3}\varphi^2 + \frac{1}{4\pi}\varphi^3 - \frac{7}{96\pi^2}\varphi^4 - \frac{8}{9}\psi^2 - \frac{\varphi\psi^2}{\pi} - \frac{7(\varphi\psi)^2}{36\pi^2} - \frac{7\psi^4}{54\pi^2} \quad \ldots (15)$$

where $M(\varphi,\psi)$ is the weight function (Haar measure) given by:

$$M(\varphi,\psi) = \left(\sin\frac{1}{2}\left(\psi + \frac{\varphi}{2}\right)\sin\frac{1}{2}\left(\psi - \frac{\varphi}{2}\right)\sin\left(\frac{\varphi}{2}\right)\right)^2 \quad \ldots (16)$$

The colorless partition function for a quark-gluon plasma contained in a volume $V_{QGP}$ then becomes:

$$Z^{CS}(T,V_{QGP}) = \frac{8}{3\pi^2}Exp^{-\frac{B}{T}V_{QGP}}\int\int_{-\pi}^{\pi}d\left(\frac{\varphi}{2}\right)d\left(\frac{\psi}{3}\right)M(\varphi,\psi)\, Exp^{V_{QGP}T^3\, g(\varphi,\psi)} \quad \ldots (17)$$

Finally, total partition function is given by:

$$Z^{CS}(T,V) = \left[Z_{Q\bar{Q}}^{CS} Z_{GG}^{CS} Z_{\pi}^{CS} Z_{VA}^{CS}\right](T,V) = \frac{4}{9\pi^2}\int\int_{-\pi}^{\pi}d\varphi d\psi\, M(\varphi,\psi)\, Exp^{VT^3(1-h)\, G} \quad \ldots (18)$$

where $G = \left[g(\varphi,\psi) - \frac{\pi^2}{30} - \frac{B}{T^4}\right]$.

**THERMODYNAMICS WITHOUT COLORLESSNESS CONDITION:**

Let us now derive the final expressions of the order parameter, energy density and the pressure in the case without colorlessness condition:
- Order parameter:

$$\langle h^{NC}(T,V)\rangle = -\frac{1}{VT^3\left(-\frac{17}{45}\pi^2 + \frac{B}{T^4}\right)} + \frac{Exp^{VT^3\left(-\frac{17}{45}\pi^2 + \frac{B}{T^4}\right)}}{\left[Exp^{VT^3\left(-\frac{17}{45}\pi^2 + \frac{B}{T^4}\right)} - 1\right]} \ldots (19)$$

- The energy density and the pressure [11-12] are calculated using their standard definitions,

$$\varepsilon(T,V) = \frac{T^2}{V}\frac{\partial}{\partial T}[\ln Z(T,V)] \text{ and } p(T,V) = T\frac{\partial}{\partial V}[\ln Z(T,V)] \ldots (20)$$

After some calculations, we find in this case, that the energy density $\frac{\varepsilon^{NC}(T,V)}{T^4}$ and the pressure, $\frac{p^{NC}(T,V)}{T^4}$ are given by,

$$\frac{\varepsilon^{NC}(T,V)}{T^4} = \left[\frac{37}{30}\pi^2 + \frac{B}{T^4}\right] - \langle h^{NC}(T,V)\rangle\left(\frac{17}{15}\pi^2 + \frac{B}{T^4}\right) \ldots (21)$$

$$\frac{p^{NC}(T,V)}{T^4} = \left[\frac{37}{90}\pi^2 - \frac{B}{T^4}\right] - \langle h^{NC}(T,V)\rangle\left(\frac{17}{45}\pi^2 - \frac{B}{T^4}\right) \ldots (22)$$

**THERMODYNAMICS WITH COLORLESSNESS CONDITION:**

Putting in the colorlessness condition, we have
- Order parameter:

$$\langle h^{CS}(T,V)\rangle = \frac{\int_0^1 h \, Z^{CS}(T,V)dh}{\int_0^1 Z^{CS}(T,V)dh} = \frac{\int_0^1 h \left[\iint_{-\pi}^{\pi} d\varphi d\psi \, M(\varphi,\psi) \, Exp^{VT^3(1-h) \, G}\right]dh}{\int_0^1 \left[\iint_{-\pi}^{\pi} d\varphi d\psi \, M(\varphi,\psi) \, Exp^{VT^3(1-h) \, G}\right]dh} \ldots (23)$$

- With $a_{HG} = \frac{\pi^2}{30}$ and $\Omega = \left[3g(\varphi,\psi) - 3a_{HG} + \frac{B}{T^4}\right]$, from the definition of the energy density, we arrive to the final expression

$$\frac{\varepsilon^{CS}(T,V)}{T^4} = \left(\frac{\pi^2}{10} + \frac{1}{\int_0^1 dh \, Z^{CS}(T,V)}\int_0^1 dh \, (-h)\left[\iint_{-\pi}^{\pi} d\varphi d\psi \, M(\varphi,\psi)\Omega Exp^{(1-h) \, T^3V \, G}\right]\right) \ldots (24)$$

The pressure of system is given by:

$$\frac{p^{CS}(T,V)}{T^4} = \left(\frac{\pi^2}{30} + \frac{1}{\int_0^1 dh \, Z^{CS}(T,V)}\int_0^1 dh \, (-h)\left[\iint_{-\pi}^{\pi} d\varphi d\psi \, M(\varphi,\psi)\boldsymbol{G} Exp^{(1-h) \, T^3V \, G}\right]\right) \ldots (25)$$

A numerical calculation is performed using Monte Carlo method by Mathematica software [13-14], to compute some thermodynamics quantities like the order parameter $h(T,V)$, energy density $\varepsilon(T,V)$ and pressure $p(T,V)$ in this work.

## RESULTS AND DISCUSSION

Our results have shown the effect of colorlessness condition on the transition between the HG and the QGP. The variations of each of order parameter, energy density and pressure normalized by $T^4$ with the temperature at different system sizes are presented in Fig1.(a), (c) and (d). We observed in the plots without colorlessness condition the transition points are not shifted with different volumes for the thermodynamic quantities in the system. The transition point $T_0$ is same for all sizes. This is due to no color singlet is taken into account.

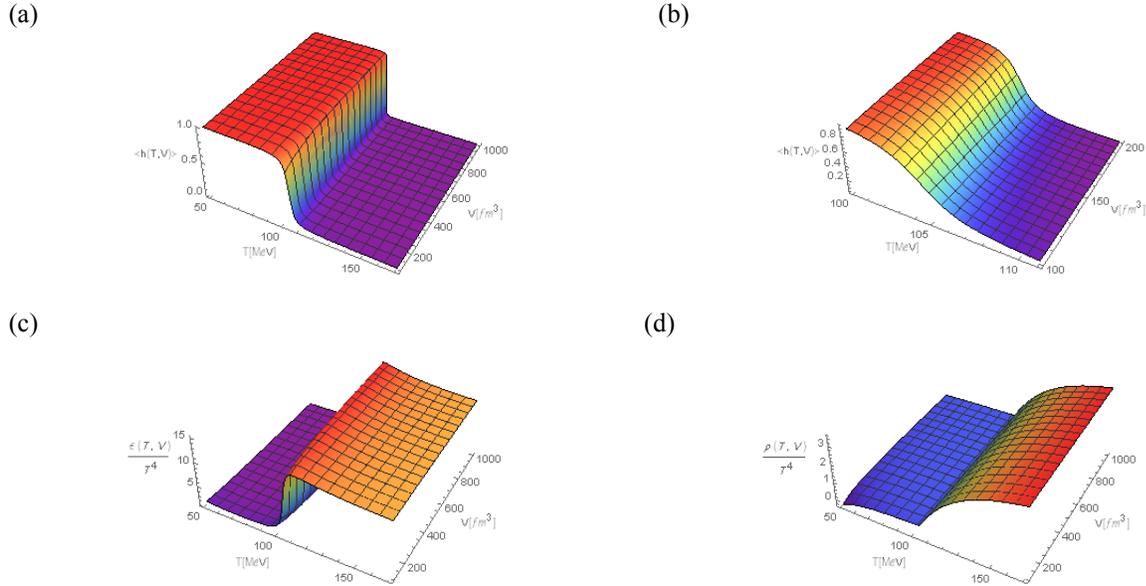

Figure 1: Some response function of the system without colorlessness condition with different volumes.
(a) and (b) order parameter (c) energy density (d) Pressure.

Meanwhile, we see the shift is very clear in Fig.2 when we include the colorlessness condition. We noticed, that the effect of the colorlessness condition disappears when approaching the thermodynamical limit. In Fig2.(a), (c) and (d) a pronounced size dependence of these quantities can be noted over almost the entire temperature range. The main and striking difference, comparatively with colorlessness condition is the shift of the transition temperature to higher values for small sizes. This is due to the effect of the colorlessness requirement which was found to lead to a gradual freezing of the effective number of degrees of freedom in the QGP [10] and [15]. In a finite volume, the pressure at a given temperature has a lower value and the equilibrium between the two phases according to the Gibbs criterion, is then reached at temperatures greater than $T_0 (\infty)$. This freezing in the QGP degrees of freedom can be more clearly seen in Figs (2c and 2d) where we compare the normalized energy density $\frac{\varepsilon(T,V)}{T^4}$.

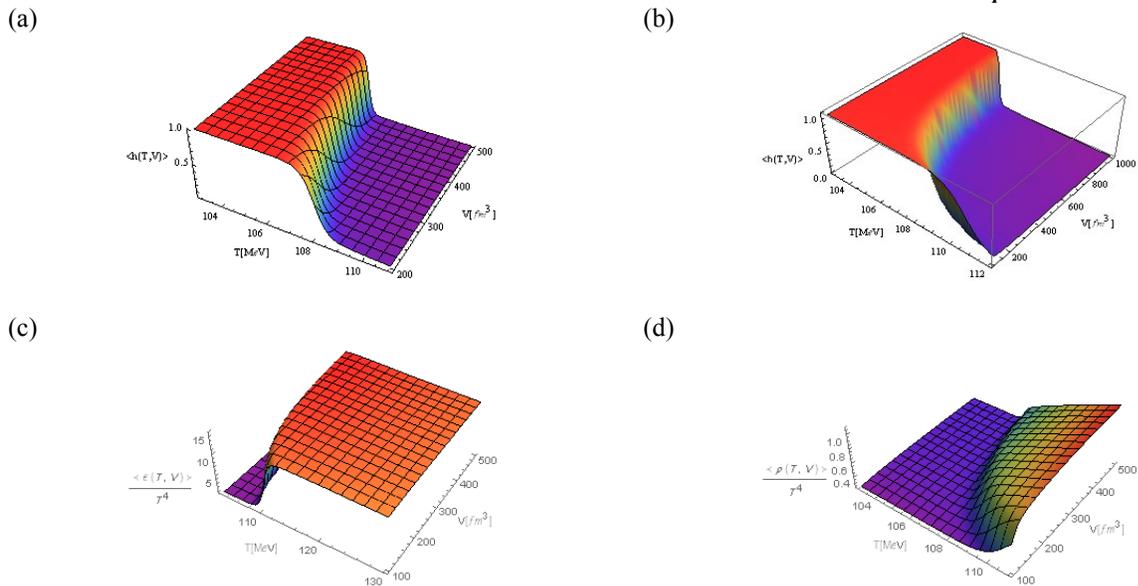

Figure 2: Some response function of the system colorlessness condition with different volumes.
(a) and (b) order parameter (c) energy density (d) Pressure.

This can better be seen in Fig.(2c) where the normalized energy density vs temperature at different volumes is illustrated more clearly. The calculated pressure in units of the fourth power of the temperature in the entire range of temperatures is shown in Fig.(2d). A rapid growth of the pressure is clearly seen just after the transition temperature, especially when the volume approaches the thermodynamical limit.

**CONCLUSION**

The work has clarified the role of the finite volume of the system and the colorlessness condition on the behavior of some response functions in the vicinity of the transition point during the deconfinement phase transition from HG to QGP. The colorlessness condition emerges naturally from the confinement phenomenon of the color charge in QCD. For this, we used the projection method given by Redlich and Turko in [9] under the internal symmetry group to take into account only the colorless states in the calculation of the partition function, so the shifting of the transition point was observed at finite volume. The sharp transition which is observed in the thermodynamical limit in some thermodynamical quantities like the order parameter, energy density and pressure at transition temperature $T_0(\infty)$, is rounded in finite volumes, and the variations of those thermodynamical quantities are perfectly smoothed on the limited range of temperature. Moreover, without this condition, the finite volume effect on the position of the transition point (shifting) is unobservable.